# Post-prognostics decision in Cyber-Physical Systems

Safa MERAGHNI, Labib Sadek TERRISSA, Soheyb AYAD
LINFI Laboratory
University of Biskra
Biskra, Algeria
meraghni.safa@gmail.com,
terrissalabib@gmail.com,
ayad_soheyb@yahoo.fr

Noureddine ZERHOUNI, Christophe VARNIER
Institut FEMTO-ST
UMR CNRS 6174 - UFC / ENSMM / UTBM
(AS2M)
Besançon, France
zerhouni@ens2m.fr,
christophe.varnier@ens2m.fr

*Abstract*— **Prognostics and Health Management (PHM) offers several benefits for predictive maintenance. It predicts the future behavior of a system as well as its Remaining Useful Life (RUL). This RUL is used to planned the maintenance operation to avoid the failure, the stop time and optimize the cost of the maintenance and failure. However, with the development of the industry the assets are nowadays distributed this is why the PHM needs to be developed using the new IT. In our work we propose a PHM solution based on Cyber physical system where the physical side is connected to the analyze process of the PHM which are developed in the cloud to be shared and to benefit of the cloud characteristics**

*Keywords— Cyber physical systems CPS, Prognostics Health Management PHM, Decision post-prognostics, cloud computing, Internet of Things.*

I. INTRODUCTION

The CPS one of the most significant advanced technologies, they connect the physical word with the cyber word using a communication layout. The physical word is the assets which are connected to sensors and data is sent to cyber word. In the cyber word, the data collected is processed [6]. Recently CPS has being used in many area such as manufacturing, monitoring and maintenance[7][8].

The industry looks to increase the available of the asset at the same time minimizing the cost of maintenance. With the development of the industry and the use of sensors The predictive maintenance is used continuously to avoid failures. The maintenance is carried out when certain indicators give the signaling that the equipment is deteriorating and the failure probability is increasing. Prognostics and Health Management (PHM) represents a great opportunity to detect upcoming failures [2]. It predict the health of an asset and its remaining useful life [3]. Using these information can be used to planned efficiently the maintenance action which can impact the reliability, the availability, the security and the quality of the final products [4].

In this work we propose a PHM solution based on CPS. We based on the CPS model and we present the PHM in three layout, the physical word where we present the assets and sensors, cyber word using the cloud we developed the PHM process for analyzing data, predicting the RUL and elaborate the planning of maintenance.

The remainder of the paper is organized as follow: In Section 2, we describe and discuss the prognostics health management and post-prognostics decision. Cloud Computing, Internet of things technologies and Cyber physical system are developed in sections 3,4 and 5. Finally the Proposed framework is developed.

II. POST-PROGNOSTICS DESICION

The Post-Prognostic Decision is one of the main steps in the Prognostics and Health Management (PHM). The PHM start by collecting data from different sensors. these data are analyzed, filtered and synchronized to be used by the prognostics step. This step predicts the health of the asset and its remaining useful life (RUL) which provide an early warning of failure. This early warning is exploited in the post prognostics step to forecast planned-maintenance and avoid unanticipated operational problems[2].

Decision Post-prognostics process use the information from prognostics as the remaining useful life to make the optimal

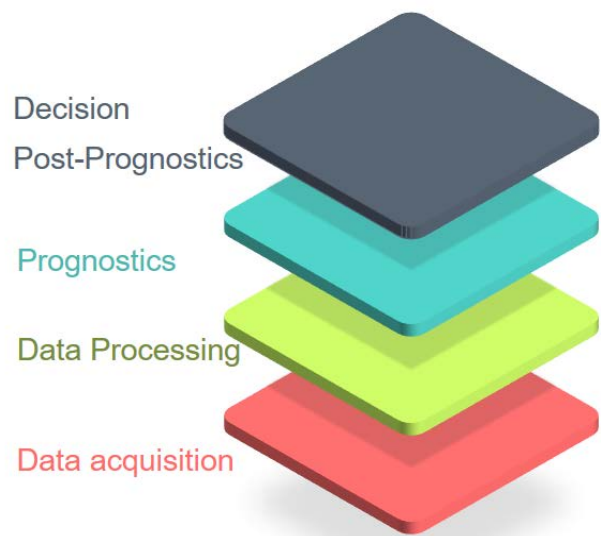

Figure1 : PHM architecture

decision of what maintenance, when and who must do this operations [5] [9].

A few works was interested to post prognostics decision, Chretien and al. [9] propose a post-prognostics decision approach to optimize the commitment of Fuel Cell Systems. And Iyrs and al. [5] developed a decision support system using decision post-prognostics.

### III. PHM SOLUTION REQUIREMENTS

The prognostics and post-prognostics decision results didn't depend just of the choice of the methods or how data is processed. But also, of many other point as:

- An overall view of all the asset.
- Storage data of the large number of information from different[14].
- Collaboration between maintenance team, Therefore, the experience and knowledge are shared [15].
- Knowledge management to be used in future cases and in the estimation of the RUL[16].
- Methods: it is important to develop robust algorithm for the estimation of the RUL and for the decision making with a high rate of confidence [16].
- Time of response from data acquisition to decision making must be reduced [17].
- Computation resource are required to process collected date in real time [17].

### IV. CYBER PHYSICAL SYSTEMS

The Cyber-physical systems (CPS) is the integration of the physical space (equipment, devices and human) with computation, communication and control systems (cyber space) [18]. The National Institute of Standards and Terminology (NIST) define the CPS as *"Cyber-physical systems (CPS) are engineered systems that are built from, and depend upon, the seamless integration of computational algorithms and physical components. Advances in CPS will enable capability, adaptability, scalability, resiliency, safety, security, and usability that will far exceed the simple embedded systems of today. CPS technology will transform the way people interact with engineered system"*.

Nowadays, Cyber- physical systems have been in various sectors such as automotive, aerospace, civil, railways, medical and manufacturing. Many works developed a PHM based CPS. [18] discussed the application of cyber-physical systems in the future of maintenance strategies to develop a smart maintenance strategy. For developing PHM for complex machinery and processes [19] focuses on existing trends in the development of industrial big data analytics and CPS.

With the advances in wireless sensors, mobile computation and big data, a lot of works integrate CPS with cloud computing [20][21][22].

### V. CLOUD COMPUTING

Cloud computing can provide a powerful, secure and easy way to storage massive data and processing infrastructure to perform both online and offline analysis and mining of the heterogeneous sensor data streams [25]. In cloud computing distributed resources are encapsulated into cloud services, SaaS (Software as a Service), PaaS (Platform as a Service) and IaaS (Infrastructure as a Service) [23]. Clients can use cloud services according to their requirements. Cloud users can request services ranging from product design, manufacturing, testing, management, and all other stages of a product life cycle [24].

#### A. Cloud benefits

There are several benefits of cloud for the maintenance process this is why many research are interested on implementation a PHM based cloud solution as Lee and al. [26] implement a PHM cloud-based platforms including developed models to real-world applications to serve the needs of industry. Yang and al. [25] propose a cloud-based prognostics system for providing a low-cost, easy-to-deploy solution for industrial big data collected in factory floors. In order to address new design requirements or resolve potential weaknesses of the original design Xia and al. [14] developed framework for the closed-loop design evolution of engineering system is proposed through the use of a machine condition monitoring system assisted by IoT and CC. Hossain and al. [27] presents a Health IoT-enabled monitoring framework, where ECG and other healthcare data are collected by mobile devices and sensors and securely sent to the cloud for seamless access by healthcare professionals.

The main benefits of cloud are

- Reducing the cost of managing and maintaining the IT system is one of the most important benefits of cloud computing. the concept pay-as-you-go and the elastic computing capabilities provided by cloud may save IT costs by using the resources of the cloud computing provider [28].

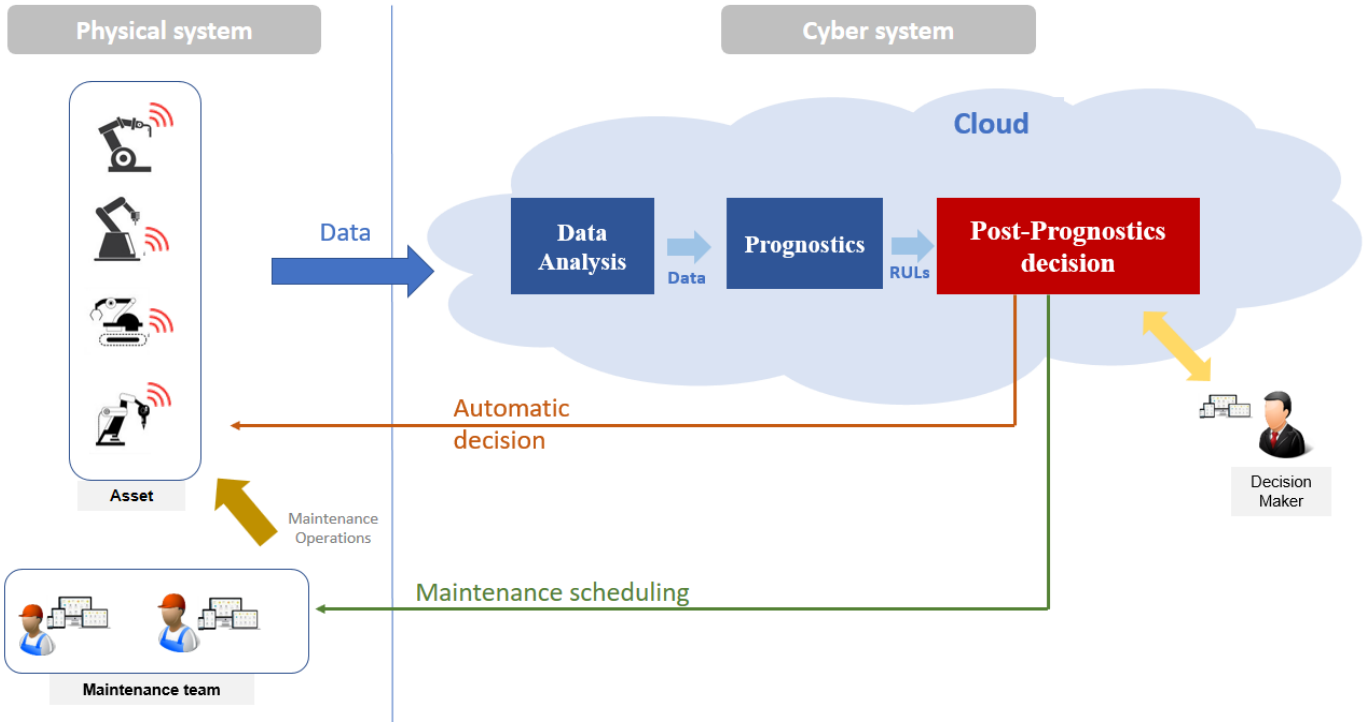

**Figure 2. Main architecture**

- Reducing the time spent to install and configured maintenance solution. Offering to users a PHM solution as a service which is already installed and configured and in few hours, they can have the application ready to use.
- The simultaneous access of resources, cloud can unify geographically dispersed resources for smart facilities management and making them available on demand to a user.
- As the cloud is composed of multiple servers and data storage units, failure of a piece of hardware does not prevent the use of the software and data.

## VI. OVERVIEW OF PROPOSED ARCHITECTURE

Figure 1 shows the frameworks architecture of the Post-prognostics decision using IoT and CC based on Cyber physical system. In the first layer present the physical space all the equipment and the maintenance team. The data from sensors are collected using the Iot. IoT will play a major role in collect, sort, synchronize and organize data in real time from different sites. In order to improve maintenance efficiency, this data will then have to be integrated in the PHM processes. The prognostics and post-prognostics process are placed in the cloud platform, using the cloud computing provide storage and computing resources in the cloud infrastructure layer to analyze and storage the huge number of data brings by IoT and sharing the decision and information provided by the framework.

the status of the system machine is analyzed through the prognostics process by the estimation of the remaining useful life. Then the post-prognostics decision according to the RUL of machine elaborates the decision of the maintenance operation that must be done.

The RUL determines broken up time of each machine, therefore the maintenance that must be applied first. But there are several constraints to make the decision of what maintenance should be done first and elaborate a maintenance scheduling. In this work, we have two constraints the cost of the labor wage and the cost of the component that must be changed.

### 1.1 Cloud-PHM layout

PHM as service can bring a lot of benefits for the maintenance results and management, the deployment of the PHM in cloud could improve efficiency of maintenance and minimizing cost and time. PHM can be provided as SaaS, PaaS or IaaS as shown in figure 2.

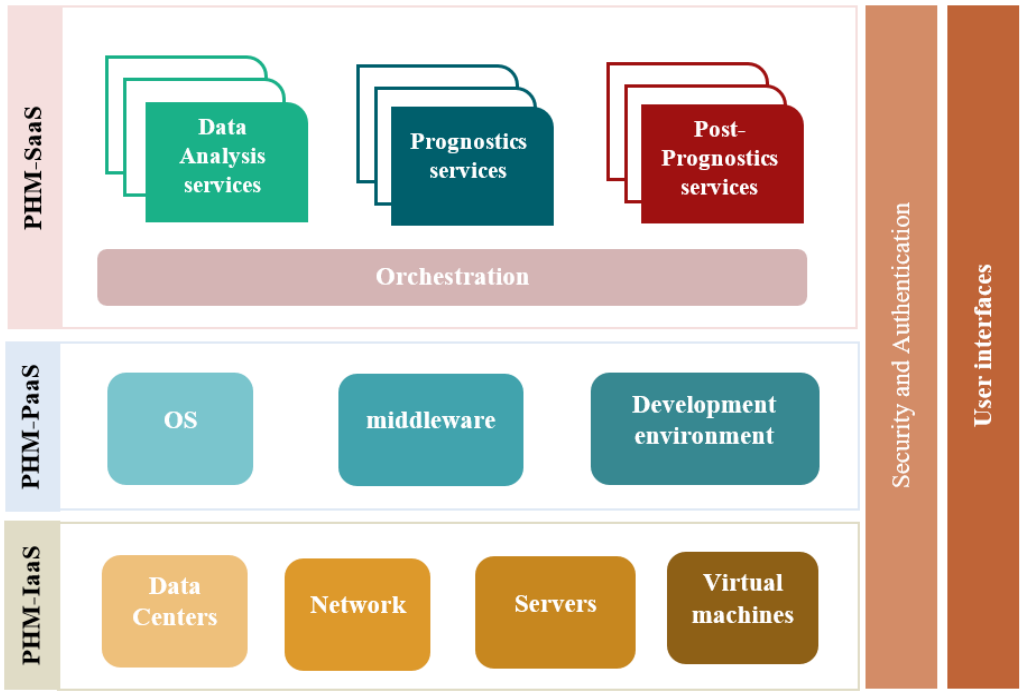

Figure3. Services of the proposed PHM based on the cloud

- PHM-SaaS: the applications are already installed and configured, this reduce time of deployment and installation, in couple of hours the application is ready to use. In each step of the PHM process There are a lot of methods which could be applied, this is why in our solution we propose this methods as services, the orchestration module synchronize and manage the connection between the PHM steps.

- PHM-PaaS: the OS, middleware and development environment are provided as services for developers and researchers, this is useful for the collaborations between persons who work in the same project. This platform allowed to change the existing services or development of new one which are used in the PHM process.

- PHM-IaaS: the infrastructure necessary for the deployment of the PHM solution is provided as services as PaaS for developers and researchers who want to test and manage their infrastructure.

## VII. CASE STUDY

### A. Problem statement

In this paper, we are interested on post-prognostics decision services. These services represent maintenance cost optimization methods that elaborate the planning of maintenance tasks to ensure the availability and the efficiency of all WTs with the optimal cost.

The maintenance cost depends on various parameters: spare parts cost, maintenance labors (manpower), subcontracting, traveling cost, …etc. each Post-Prognostics decision service can optimize one or many of these costs and can use one method or more.

In this work we propose a post-prognostics service to optimize the traveling cost. this cost depends on the distance crossed by the maintenance technicians. To minimize the traveling cost we minimized the distance traveled by maintenance team using genetic algorithm.

The assets are distributed in many sites. For each asset we have its position and its RUL provided by the prognostics service. and we have one maintenance center where all the maintenance team is. The maintenance team start the travel from the maintenance center and should visit all assets before the failure. There for, we are looking for the shortest distance traveled by maintenance team starting from the maintenance operation that passed by all assets before their RUL.

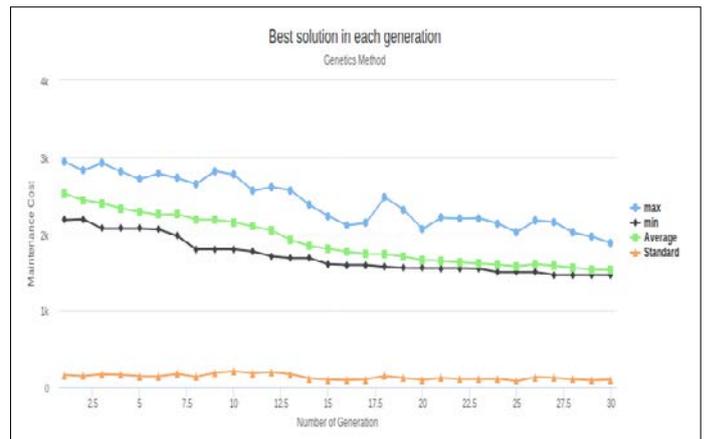

Fig.4. Best solution for 30 generation with population size =100

GA are stochastic-based search techniques particularly suitable for solving complex optimization problems [29], GA are applied to maintenance optimization because of their robust search capabilities that resolve the computational complexity of large-size optimization problems [30]. In [31] GA are used, as an optimization tool to compare the cost of premature replacement with the cost of downtime if grounded for the sole purpose of replacement. [32] show the application of GA to optimize the sub-station LCC by finding the best maintenance strategy. A GA that optimizes system availability, and cost with system-maintenance constraints is presented in [29] .

In our work the genetic algorithm is used to optimize the distance traveled by maintenance team. We start by random population and apply uniform partially matched crossover, uniform mutation and tournament selection. This method was developed in Python using Deap1.1.0 library.

Figure 4 show the evolution of the algorithm. Th black curve present the minimum cost with is the optimal one

## VIII.  Conclusion

The application of PHM enhances the effective reliability and availability of a system in its life-cycle conditions by detecting upcoming failures and reducing the unscheduled maintenance. The prognostics predicted the remaining useful life (RUL) and the decision post-prognostics process planned maintenance activities according to these RULs.
The industry system in our days are more complex and distributed in different sites, this need more resources to storage data and availability in a distributed environment. Collaboration, Internet of things and cloud has been used in this paper to collect information from machine, analyze and storage this data to provide a decision post prognostics solution as a cloud service.
Genetic algorithms are employed to reduce the cost of malignance labors and elaborate the planning of maintenance operations.